\documentclass[aps,prB,preprint,superscriptaddress]{revtex4-1} 
\usepackage{amssymb}
\usepackage{amsmath}
\usepackage{graphics}

\ifx\pdfoutput\undefined

\usepackage[dvips]{graphicx}

\else

\usepackage[pdftex]{graphicx}

\fi
\usepackage{epstopdf}
\newcommand {\bfx} {\mathbf{x}}
\newcommand {\dd} {\mathrm{d}}

\newcommand {\kB} {k_{\mathrm{B}}}

\newif\ifRMHC
\RMHCtrue
\RMHCfalse

\newif\ifAJHC
\AJHCtrue
\AJHCfalse

\begin{document}
\title{New Methods for Calculating the Free Energy of Charged Defects in Solid Electrolytes}


\author{Robert M. Horton}
\thanks{Corresponding Author: \texttt{rh609@imperial.ac.uk}}
\affiliation{Department of Physics, Imperial College, London, SW7 2AZ}
\affiliation{Department of Materials, Imperial College, London, SW7 2AZ}
\author{Andrew J. Haslam}
\affiliation{Department of Chemical Engineering, Imperial College, London, SW7 2AZ}
\author{Amparo Galindo}
\affiliation{Department of Chemical Engineering, Imperial College, London, SW7 2AZ}
\author{George Jackson}
\affiliation{Department of Chemical Engineering, Imperial College, London, SW7 2AZ}
\author{Michael W. Finnis}
\affiliation{Department of Physics, Imperial College, London, SW7 2AZ}
\affiliation{Department of Materials, Imperial College, London, SW7 2AZ}

\date{\today}


\begin{abstract}
A methodology for calculating the contribution of charged defects to the configurational free energy of an ionic crystal is introduced. The temperature-independent Wang-Landau Monte Carlo technique is applied to a simple model of a solid electrolyte,
consisting of charged positive and negative defects on a lattice. The electrostatic energy is computed on lattices with periodic boundary conditions,  
and used to calculate the density of states and statistical-thermodynamic potentials of this system. 
The free energy as a function of defect concentration and temperature
  is accurately described by  a regular solution model up to concentrations of $10\%$ of defects, well beyond the range described by the ideal solution theory. 
  The approach, supplemented by short-ranged terms in the energy, is proposed  as an alternative to free-energy methods that require a number of simulations to be carried out over a range of temperatures.
\end{abstract}

\maketitle


 
\section{Introduction}
The presence of charged defects gives rise to useful electrolytic behaviour in many materials of practical interest.
In solid electrolytes,  such as yttria-stabilized zirconia (YSZ), vacancies are present in concentrations well above 1\%. 
(This is in contrast to simple oxides where the vacancy concentrations are usually very low.)
 This high concentration of (locally charged) vacancies is essential in making these materials useful in applications; lattice diffusion of highly concentrated charged vacancies is primarily responsible for the useful electrolytic behaviour in such materials \cite{Goff1999,Owens2000}.

More generally, the behaviour of solid electrolytes of this type is determined by gradients in electrochemical potential, which provide the driving forces for point-defect diffusion. The chemical potentials in turn depend on the free energy as a function of defect concentrations.
Accordingly, for modelling such systems, \textit{e.g.}, using phase-field methods, or other implementations of irreversible thermodynamics, a detailed understanding of the thermodynamic behaviour of these defects is highly desirable.

A major challenge to the modelling of the thermodynamics of solid-electrolyte systems is the long-ranged nature of the Coulombic defect-defect interactions. For example, the long-ranged character of the Coulombic interactions significantly affects defect mobilities.\cite{Wang1981,Inaba1996}
However, It is unclear how best to incorporate these interactions in an empirical description of the free energy as a function of concentrations, such as that used in the Calphad methodology \cite{Saunders1998,Lukas2007}.
We propose an approach that overcomes this challenge through the use of an efficient atomistic-simulation method to characterise the thermodynamic functions.

We use an implementation of Wang-Landau (WL) \cite{Wang2001,Wang2001a} sampling, a biased Monte Carlo method, 
in a lattice-based simulation to obtain the Helmholtz free energy of the system at any temperature as a function of concentration. This free energy can then be used to parameterise macroscopic thermodynamic approaches, such as the subregular solution (see, \textit{e.g.}, \cite{Lukas2007}) with Redlich-Kister \cite{Redlich1948} polynomials. 

We choose the WL method as it is a sampling method that produces a temperature-independent density of states for the system within a single simulation. 
This allows the partition function, and hence all the statistical-thermodynamic properties of the system, to be evaluated as functions of temperature using simple quadratures. 

This paper is organised in as follows. Our model system based on YSZ is described in section \ref{theModel}. We outline our methodology,
including our implementation of the WL method in section \ref{sec:Methodology}. Section \ref{sec:Results} reports  the results of applying this methodology to the model system  described in section, most importantly  we describe and analyse our calculations of the Helmholtz free energy of the system as a function of temperature and defect concentration. This free energy is compared to that obtained using the ideal-solution theory (see {\textit{e.g.}}, \cite{Callen1985}), and to that from Debye-H\"{u}ckel theory \cite{Debye1923}. As a proof of principle for future applications, we show that the excess free energy obtained from simulation is accurately represented as a function of defect concentration by low order polynomials with temperature dependent coefficients, in the manner familiar to practitioners of phase diagram assessment in CALPHAD \cite{Saunders1998,Lukas2007}.   Finally, our conclusions are presented in section \ref{sec:Conclusions}.

\section{The Model}
\label{theModel}
The high-temperature fluorite structure of YSZ is chosen as the basis for our simple model as a prototypical example of a material in which charged defects not only exist in high concentrations, but also play a principal role in its use as an ionic conductor. 
When zirconia ($\mathrm{Z}\mathrm{O}_{2}$) is doped with yttria ($\mathrm{Y}_{2}\mathrm{O}_{3}$) the yttrium ions ($\mathrm{Y}^{3+}$) replace zirconium ions ($\mathrm{Zr}^{4+}$) on a face-centred cubic (FCC) lattice. As there is a  difference in the net charge of these ions, charge neutrality in the system is preserved by the  creation of  locally-charged oxygen vacancies on a simple-cubic (SC) sub-lattice nested within the FCC lattice. Substitutional  yttrium ions and oxygen vacancies are assumed to carry local excess charges of -1 and +2 respectively with respect to the perfect  lattice, so to preserve overall charge neutrality the ration of oxygen vacancies to yttrium ions should be 1:2. The high concentration and mobility of these $+2$ charged oxygen vacancies  gives YSZ its useful electrolytic properties.

We assume that the yttrium substitutional ions and oxygen vacancies behave as $-1$ and $+2$ point charges and Coulomb's law is treated with periodic boundary conditions by  Ewald summation (see, \textit{e.g.}, \cite{Frenkel2002}). 
The volume, the number of lattice sites and the numbers of charges in each simulation of the  electrolyte remains constant.
The only length scale in this model system is the lattice parameter. For convenience we use the physical value of $0.51$\,nm, appropriate to YSZ \cite{Ingel1986}. 
In the present case its role, along with a dielectric constant of $30$ (relative to vacuum) \cite{Samara1990}, is to  scale the energies; this could equally well be done after the simulation, however, it is convenient to  generate energies and temperatures directly in the familiar units. Moreover, when a more-realistic potential is used to augment the Coulomb potential, in order to take into account the short-ranged interactions, the correct choice of lattice parameter is essential. 
\begin{figure}
\center
\includegraphics[width=0.4\columnwidth]{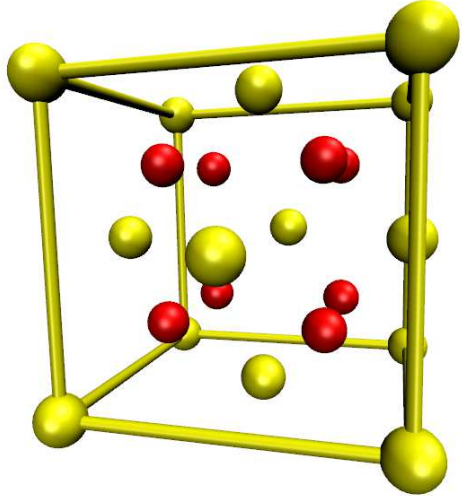}
\caption{The fluorite lattice composed of a SC sub-lattice (Red) nested within a FCC lattice (yellow).}
\label{unitcell}
\end{figure}
Unless otherwise stated, all simulations are carried out within supercells of $4\times4\times4$ fluorite unit cells with periodic boundary conditions one unit cell of which is depicted in Figure \ref{unitcell}.

\section{Methodology}
\label{sec:Methodology}
In classical statistical mechanics we are interested in evaluating configurational integrals or partition functions $\mathcal{Z}$ of the form
\begin{equation}
\mathcal{Z} = \!\!\int\!\!\exp(-\beta E(x_1,x_2,\cdots,x_{3N}))\dd x_1\cdots\dd x_{3N}.
\label{eqn:PS1}
\end{equation}
These may be written more concisely as
\begin{equation}
\mathcal{Z} = \!\!\int\! \exp\left(-\beta E(\bfx)\right)\dd\bfx,
\end{equation}
where $\beta=\left(1/ \kB T\right)$, $\kB$ is Boltzmann's constant, $T$ is the absolute temperature, $E(\bfx)$ is the potential energy of a  microstate at $\bfx$, which represents a point in the $3N$-dimensional  space of  possible atomic positions. In a complete partition function the kinetic energy is added to the potential, with integration extending also over the  $3N$ momental;  this is done analytically to give a familiar temperature-dependent de Broglie prefactor in front of the above integral. In our model, however,  the kinetic energy is zero, and will not concern us here. 

All the thermodynamic functions can be expressed either directly as Boltzmann averages, or in terms of derivatives of $\mathcal{Z}$ with respect to $\beta$.
For example the Helmholtz free energy $\left(F\right)$ of the system is
\begin{equation}
 F=-\beta^{-1} \ln\left(\mathcal{Z}\right)
 \label{eqn:helmholtzFreeEnergy}
\end{equation}
and the internal energy $\left(U\right)$ is
\begin{eqnarray}
 U&=&-\frac{\partial\ln\left(\mathcal{Z}\right)}{\partial\beta}\\
 &=&\frac{\int\! E(\bfx)\exp\left(-\beta E(\bfx)\right)\dd\bfx}{\mathcal{Z}}.
 \label{eqn:internalEnergy}
\end{eqnarray}


Evaluating the configurational integral (\ref{eqn:PS1}) or its derivatives by a brute-force method is usually highly inefficient, because the variation in energy of a small number of the possible microstates at low energy may have a disproportionate impact on the results. 
More-sophisticated methods are therefore highly desirable. 
Traditionally, Metropolis Monte Carlo (MMC) \cite{Metropolis} has been the method of choice for evaluating thermodynamic functions of a given system.
However, MMC is carried out in the canonical ensemble and, as a consequence, a simulation is required for every temperature one wishes to examine; consequently, the approach can be computationally expensive.
The formulation of the method also means that evaluating properties that are not characterized as Boltzmann distributions, notably $F$ or the entropy $S$, is more challenging, and it is normally done by integrating a function that \emph{can} be calculated as a configurational average, which requires a whole series of  long simulations to obtain the integrand at each point on the path.   

The problems associated with MMC have meant that in recent years there has been a renewed interest in developing Monte Carlo methods that allow the full thermodynamics of a system to be calculated from a single simulation \cite{Swendsen1987,Berg1992,Lee1993}. 
One of the more-popular of these methods is an approach developed by Wang and Landau \cite{Wang2001,Wang2001a}, 
which enables the evaluation of the configurational integral through the determination of the density of states of the system. 
This method is widely used, has been generalized for off-lattice systems \cite{Shell2002}, and has now been applied to systems ranging from metals \cite{Eisenbach2009} to proteins \cite{Rathore2004}. It has also been incorporated into molecular-dynamics simulations \cite{Shimoyama2011}.

\subsection{Wang-Landau Sampling}

Within the WL formalism one proceeds from a reformulation of the configurational integral, equation (\ref{eqn:PS1}), in terms of the normalised density of states of the system $g(E)$ at energy $E$:
\begin{equation}
\mathcal{Z}=\Omega\int g\left(E\right)\mathrm{exp}\left(-\beta\; E\right)\mathrm{d}E,
\label{DOS_Partition}
\end{equation}
where 
\begin{equation}
g\left(E\right)=\Omega^{-1}\int\delta\left(E-E\left(\mathbf{x}\right)\right)\mathrm{d}\mathbf{x}
\end{equation}
is the normalised density of states, $\Omega$ is the total number of microstates of the system, and $\delta$ is the Dirac delta function.
It is apparent from Equation (\ref{DOS_Partition}) that given $g(E)$ (and knowledge of $\Omega$)  one could calculate the configurational integrals and, thus, the thermodynamics of the system at any temperature by simple quadrature. 
Unfortunately, in most cases of interest, one has no {\it{a priori}} knowledge of $g(E)$. The aim of the WL method is to sample the system in such a way that $g(E)$ can be accumulated as a single simulation progresses.
As in conventional Monte-Carlo methods, a `simulation' refers to the process of evolving the system from microstate to microstate in a series of random steps that are accepted or rejected according to specific criteria. 
Thus it does not imply any attempt to simulate a physical trajectories in phase space. 
The term `walker' is used to refer to a member of an ensemble of microstates that evolve simultaneously by steps of the walkers during the simulation.

Following \cite{Brown2011}, one first seeks to obtain a Metropolis-like acceptance criterion from the detailed-balance condition requiring that, at equilibrium, the net flux of random walkers between the set of microstates with energy $E_{i}$ and those with energy $E_{i+1}$ must be zero. This condition can be expressed in terms of the density of walkers, $n\left(E_{i}\right)$, the probability, $P\left(E_{i+1}|E_{i}\right)$, of attempting to make a move from an initial microstate among those with energy $E_{i}$ to a final microstate among those with energy $E_{i+1}$,  and the probability, $P^{\mathrm{A}}_{E_{i}\rightarrow E_{i+1}}$, that this move will be accepted:
\begin{eqnarray}
 n\left(E_{i}\right)&P&\left(E_{i+1}|E_{i}\right)P^{\mathrm{A}}_{E_{i}\rightarrow E_{i+1}}\nonumber\\ 
 &=&n\left(E_{i+1}\right)P\left(E_{i}|E_{i+1}\right)P^{\mathrm{A}}_{E_{i+1}\rightarrow E_{i}}.\;
 \label{detailedBalance}
\end{eqnarray}
We observe that if the system is sampled evenly in energy space, \textit{i.e.}, if 
\begin{equation}
 n\left(E_{j}\right)=n\left(E_{j+1}\right),
 \label{eqn:equalSamplingCondition}
\end{equation}
 one obtains
\begin{eqnarray}
\label{eqn:evenenergysampling}
 P\left(E_{i+1}|E_{i}\right)&P&^{\mathrm{A}}_{E_{i}\rightarrow E_{i+1}}=\nonumber\\
 &P&\left(E_{i}|E_{i+1}\right)P^{\mathrm{A}}_{E_{i+1}\rightarrow E_{i}}.
 \end{eqnarray}
 Conversely, if we can define an acceptance criterion that will ensure (\ref{eqn:evenenergysampling}) is satisfied, then from (\ref{detailedBalance}) the sample population $n(E_{i})$ must be uniformly distributed in energy. This is rather straightforward to achieve, as follows.

If the microstate to which a random walker will attempt to move is chosen in an unbiased manner,  the probability of attempting to move to any given microstate with energy $E$ is equal to the normalised density of states at that energy (\textit{i.e.}, $P\left(E_{j+1}|E_{j}\right)=g\left(E_{j+1}\right)$), so that
\begin{equation}
g\left(E_{i+1}\right)P^{\mathrm{A}}_{E_{i}\rightarrow E_{i+1}}=g\left(E_{i}\right)P^{\mathrm{A}}_{E_{i+1}\rightarrow E_{i}}.
\label{subDOS}
\end{equation}
By enforcing a Metropolis-style sampling criterion, the probability of accepting a move to a given microstate will be given by
\begin{equation}
 P^{\mathrm{A}}_{E_{j}\rightarrow E_{j+1}} = \mathrm{min}\left[1,\frac{w_{E_{j+1}}}{w_{E_{j}}}\right],
 \label{acceptance}
\end{equation}
where $w_{E_{j}}$ is a weight  associated with each energy. (In conventional MMC these weights  are set equal to the Boltzmann factor).
Substituting (\ref{acceptance}) into Equation (\ref{subDOS}) one obtains a  condition that the weights must satisfy:
\begin{equation}
 g\left(E_{i+1}\right)w_{E_{i+1}}=g\left(E_{i}\right)w_{E_{i}}.
 \label{wlsatisfy}
\end{equation}
This holds for all energies only if $w_{E_{i}}\propto\frac{1}{g\left(E_{i}\right)}$ or $w_{E_{i}}=0\;\forall\;i$. This result implies that Monte Carlo moves based on the acceptance criterion,
\begin{equation}
 \mathrm{P}^{A}_{E_{j}\rightarrow E_{j+1}} = \mathrm{min}\left[1,\frac{g\left(E_{j}\right)}{g\left(E_{j+1}\right)}\right],
 \label{WLacceptance}
\end{equation}
 will satisfy detailed balance in the form required, implying that the walkers are distributed evenly in energy.
 The strategy of WL is to ensure that this acceptance criterion is approached as the system evolves, so that the distribution of walkers will converge to a constant value of $n(E_{i})$ for all $i$.

In the WL method one proceeds by systematically improving an initial estimate of the density of states ($g_{0}(E)$) (usually this is chosen such that $g_{0}=1$). A random walk is taken according to (\ref{WLacceptance}) and, after every attempted move, the estimate of the density of states for the current energy of the system is increased by some multiplicative factor $f$. In practice one maintains a record of $\ln{g\left(E\right)}$ and increases it according to 
\begin{equation}
\mathrm{ln}\left(g_{j+1}\left(E_{i}\right)\right)=\mathrm{ln}\left(g_{j}\left(E_{i}\right)\right)+\mathrm{ln}\;f,
\label{update}
\end{equation}
where $j$ indicates a point along the random walk.
In theory this process should be repeated until the walkers have visited all regions of energy space ``equally", \textit{i.e.}, Equation (\ref{eqn:equalSamplingCondition}) is satisfied. This, in turn, means that Equation (\ref{eqn:evenenergysampling}) holds and, thereby, satisfying Equation (\ref{wlsatisfy}) implies that the estimate of the density of states
has converged to the true value. However, the probability of every region of energy space being visited an equal number of times is extremely small so, in practice, simulations would continue indefinitely if this convergence criterion were applied. Instead, in the original implementation of WL, energy space is divided into windows, and a histogram is updated every time a window is visited by a walker. The simulation is taken to be converged when this histogram satisfies some (user-defined) flatness criterion. A proof of convergence of this procedure has been given in \cite{Zhou2005}.

It was also suggested in \cite{Zhou2005} that, as the statistical fluctuations in the histogram are proportional to $\frac{1}{\sqrt{\ln f}}$, a more rigourous convergence criterion than this arbitrary, used-defined flatness criterion should be applied. They proposed that convergence should be deemed adequate when each histogram bin of energy space has been visited at least $\frac{1}{\sqrt{\ln f}}$ times. This is the criterion for convergence we have adopted in the present work.

Knowledge of the statistical fluctuations in the histogram, which is what we want to approach the density of states, also motivates the choice of value for the update factor $f$.
In principle one would like to use a value of $f$ which is as small as possible to minimise this error. However, such a choice renders the simulation very computationally expensive, since the walkers take more cycles to explore phase space. On the other hand, since the value of $f$ implicitly limits the accuracy of the convergence, a large value may not allow the density of states to converge with sufficient accuracy. 
In order to address this issue one repeats the WL procedure a number of times, using the density of states from the previous iteration obtained using a larger value of $f$, as the starting point for the next. 
By systematically reducing the update factor for each WL iteration one successively gains a better approximation of the density of states of the system whilst minimising the computational cost.

In practice one starts with $f=e$ (\textit{i.e.}, $\ln{f}=1$) and, once the convergence criterion has been met, this value of $f$ is reduced by an order of magnitude and the process is repeated, reducing the update factor in the same manner each time the convergence criterion is met.
This is carried out until the update factor has a minimal impact on the previous estimate of the density of states. 
It is at this point that one assumes an accurate estimate of the density of states has been obtained. 

As mentioned in \cite{Wang2001}, rather than calculating the exact density of states of the system, one in fact calculates a relative density of states; in practice one calculates the density of states of the system multiplied by a simulation-dependent constant. 
However,  one can readily obtain a normalised density of states from a WL simulation. Thus the only remaining unknown in Equation (\ref{DOS_Partition}) is the total number of microstates of the system $\left(\Omega\right)$.
As all the simulations to be carried out will take place on a lattice, $\Omega$ can be directly enumerated through combinatorics (see \textit{e.g.}, \cite{Callen1985}). Thus, having calculated $g\left(E\right)$ and $\Omega$, one can evaluate Equation (\ref{DOS_Partition}) and hence all required thermodynamic functions through simple quadrature.

\section{Results for Lattice Models of YSZ}
\label{sec:Results}
\begin{figure}[htp]
\begin{center}
\includegraphics[width=0.7\columnwidth]{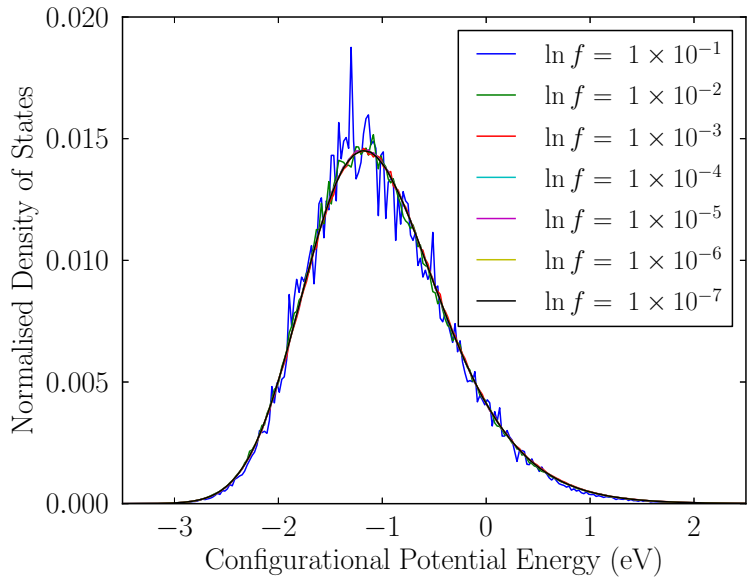} 
 \caption{Evolution of the normalised density of states of a model YSZ system consisting of 24 charged defects (16 yttrium substitutions and 8 oxygen vacancies), 
 over successive Wang-Landau iterations, with values of the update value $f$ as indicated.}
\label{convergingDOS}
\end{center}
\end{figure}

We begin by illustrating, in Figure \ref{convergingDOS}, the evolution of the estimate of $g\left(E\right)$ after each successive WL iteration for a 2:1 electrolyte system of 8 oxygen vacancies and 16 yttrium substitutions; the values of $\ln f$ that are employed are indicated in the figure. Using the large, initial value of $\ln f = 0.1$ the calculated $g\left(E\right)$ is observed to have large statistical fluctuations. However, this estimate is dramatically improved by the next iteration and, already by the third iteration, visually the estimate appears to be very good (as judged by the smoothness of the curve). The rapid convergence of the estimates illustrated in the figure demonstrates the effectiveness of the procedure.

Once calculated, $g\left(E\right)$ can be analysed by examining representative microstates over the full energy range.
\begin{figure}[htp]
\begin{center}
\includegraphics[width=0.7\columnwidth]{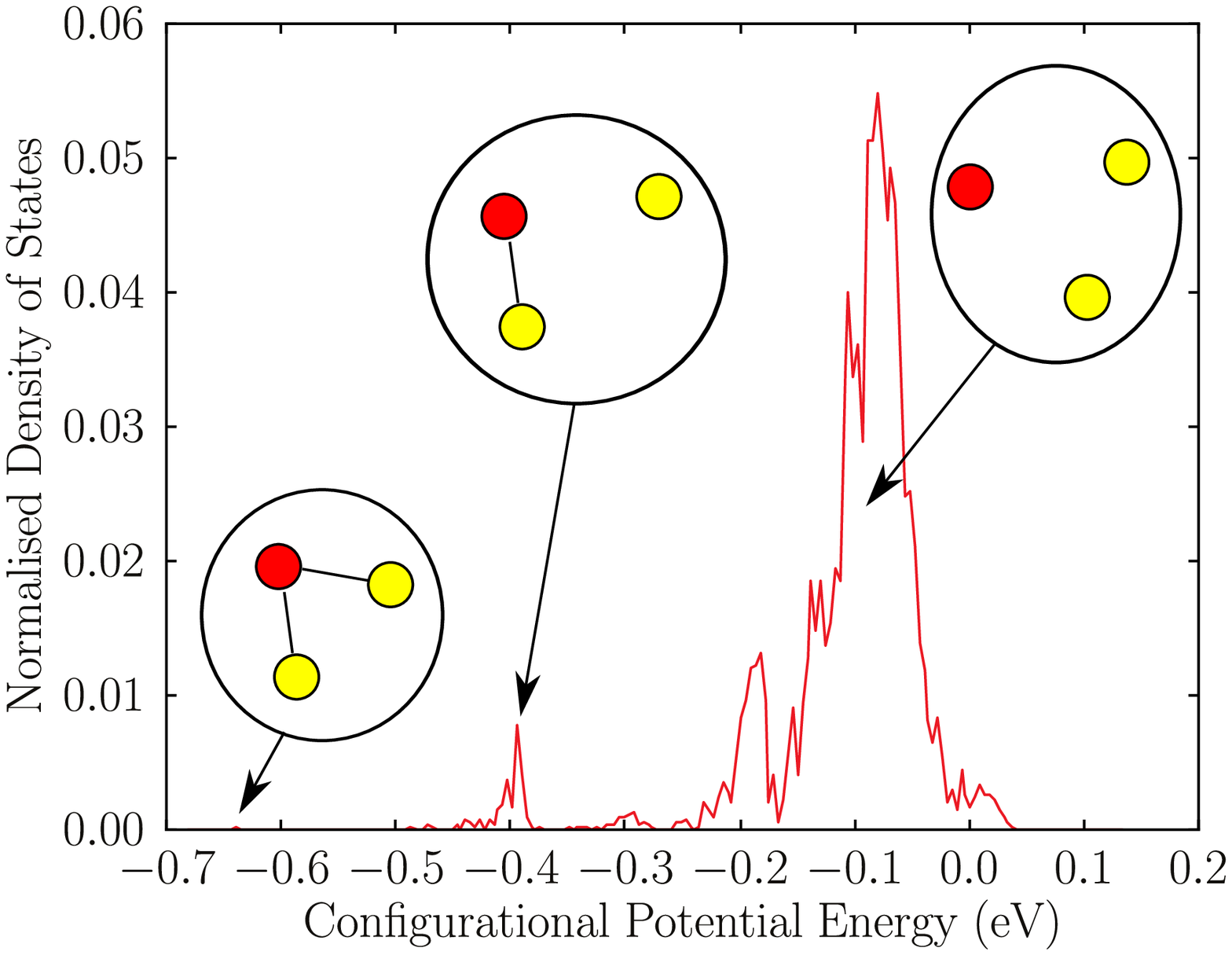} 
\caption{The normalised density of states determined from a Wang-Landau simulation approach for a system composed of one oxygen vacancy (red) and two yttrium substitutions (yellow). Characteristic microstates, and the regions of $g\left(E\right)$ in which they feature are also illustrated.}
\label{unfrozenDOS}
\end{center}
\end{figure}
Unsurprisingly, the very lowest-energy microstates of our system correspond to trimers of one oxygen vacancy and two nearest-neighbour yttrium substitutions. This is illustrated in Figure \ref{unfrozenDOS} for the simple system containing just three defects. 
By choosing a system with very few defects, the individual features in $g\left(E\right)$ are preserved and easier to discern, whereas they are ``smoothed out" when examining a system containing many defects.

\begin{figure}[htp]
\begin{center}
\includegraphics[width=0.7\columnwidth]{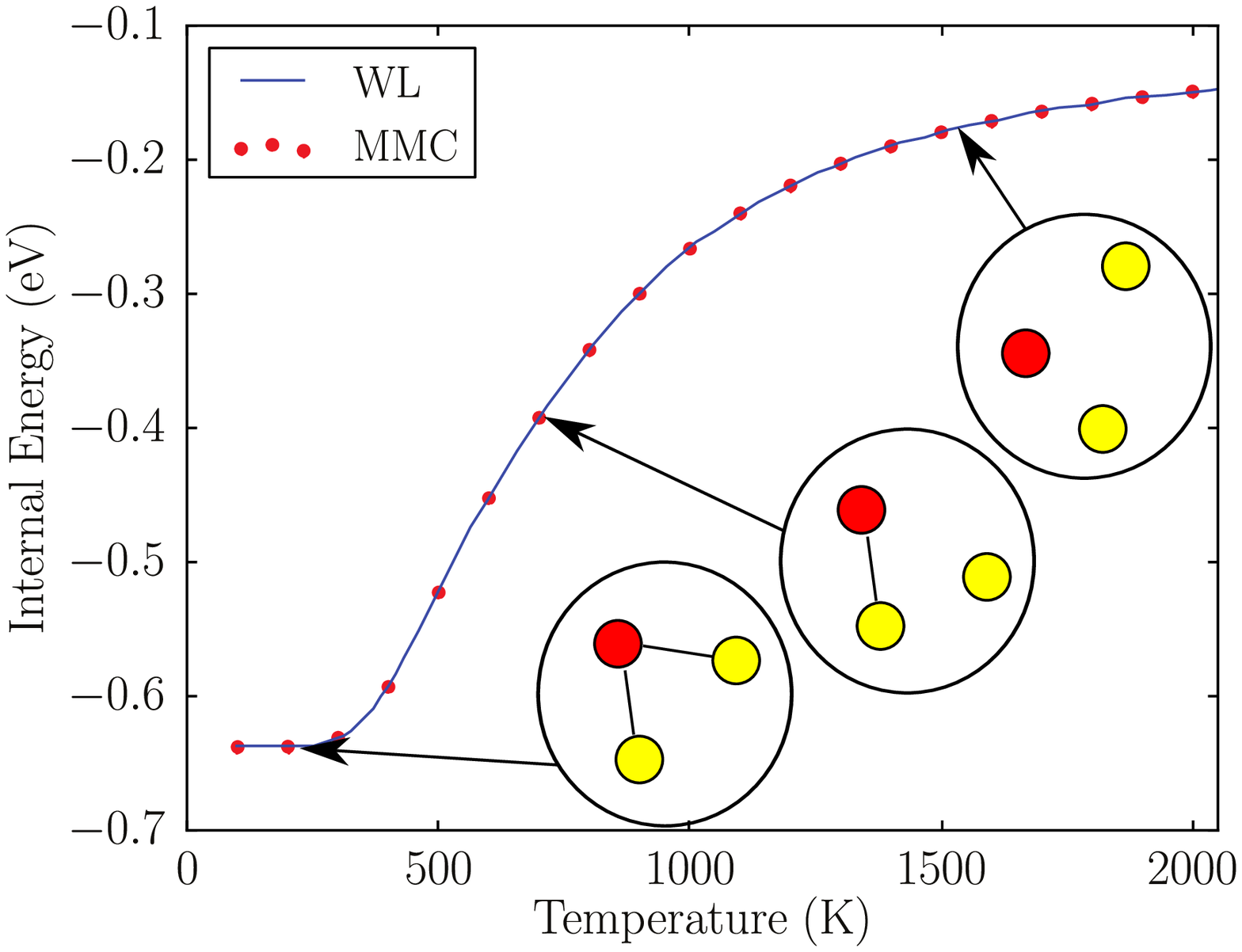} 
\caption{Temperature dependence of the configurational energy obtained with the Wang-Landau simulation approach for the model YSZ system with three charged defects (2 yttrium substitutions + 1 oxygen vacancy). Representative microstates for different temperature regimes are also indicated on the graph. }
\label{unfrozenIntEn}
\end{center}
\end{figure}

Knowledge of $g\left(E\right)$ allows one to calculate the full thermodynamics of the system by evaluating the configurational integral, equation~(\ref{DOS_Partition}), thereby obtaining the configurational internal energy, equation~(\ref{eqn:internalEnergy}) and the Helmholtz free energy, equation~(\ref{eqn:helmholtzFreeEnergy})  over the entire temperature range. In Figure \ref{unfrozenIntEn} for the internal energy of the  three-defect system we include data points corresponding to internal energies evaluated using traditional MMC simulations  in addition to the WL results; good agreement between the two methods is apparent. 
Also illustrated in the figure are the changes that can be attributed to the three different regimes the system passes through as the temperature is increased. 
As one might expect from the analysis of representative microstates at different regions of $g\left(E\right)$, with increasing energy the system evolves from a fully associated to a fully dissociated regime. 
For all sizes of system, structures composed of charge-neutral trimers are stable at lower temperatures. These dissociate successively into dimers and monomers as the temperature increases and entropic effects begin to dominate.

\begin{figure}[htp]
\begin{center}
\includegraphics[width=0.7\columnwidth]{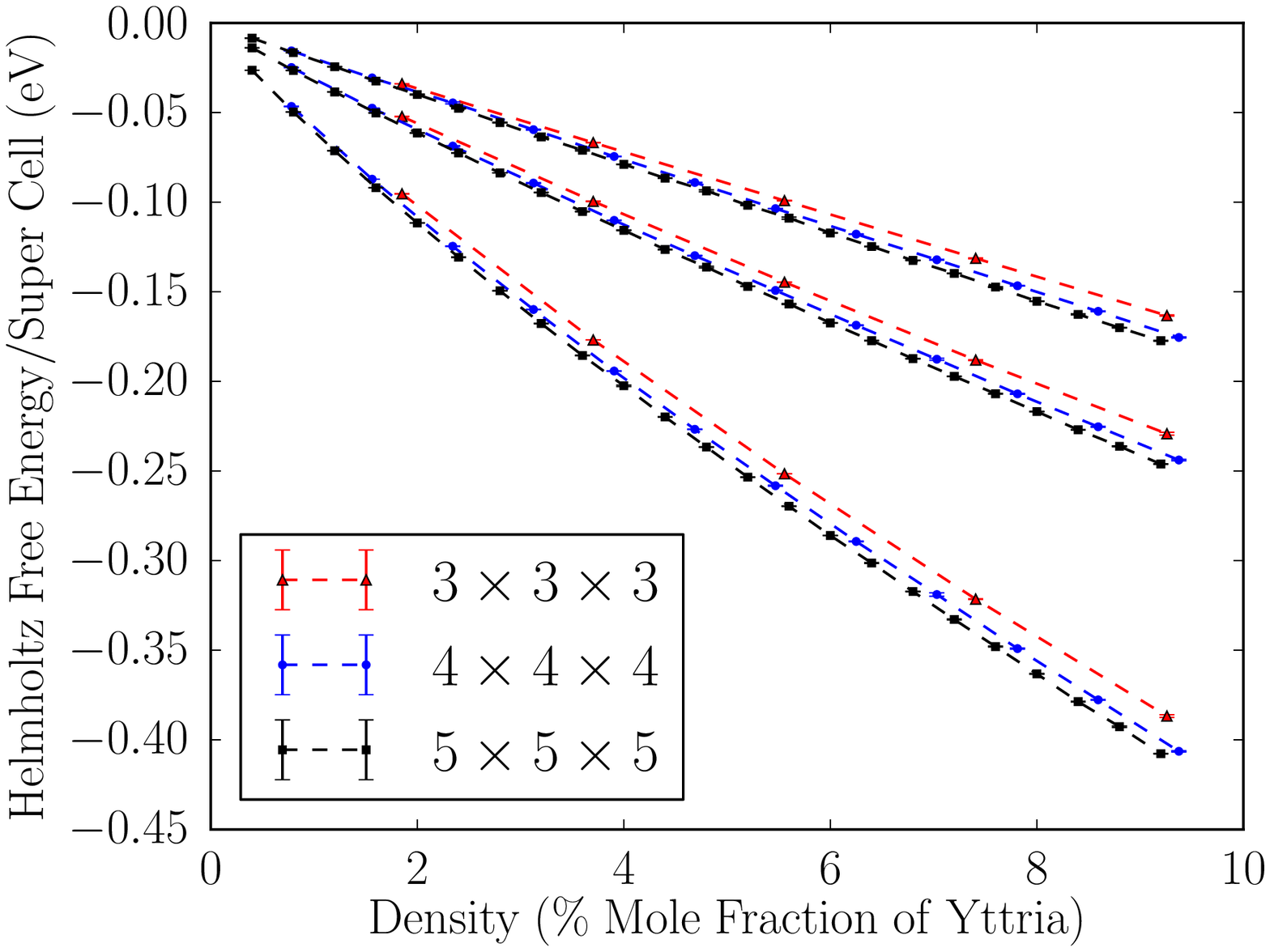} 
\caption{Comparison of the temperature and density dependence of the Helmholtz free energy of the system obtained with a Wang-Landau simulation approach for the model YSZ system as a function of system size.
The effect of system sizen on the calculated free energy is shown at temperatures of $500\,$K (top), $1000\,$K (middle) and $2000\,$K (bottom).}
\label{finiteSizeEffects}
\end{center}
\end{figure}
\begin{figure}[htp]
\begin{center}
\includegraphics[width=0.7\columnwidth]{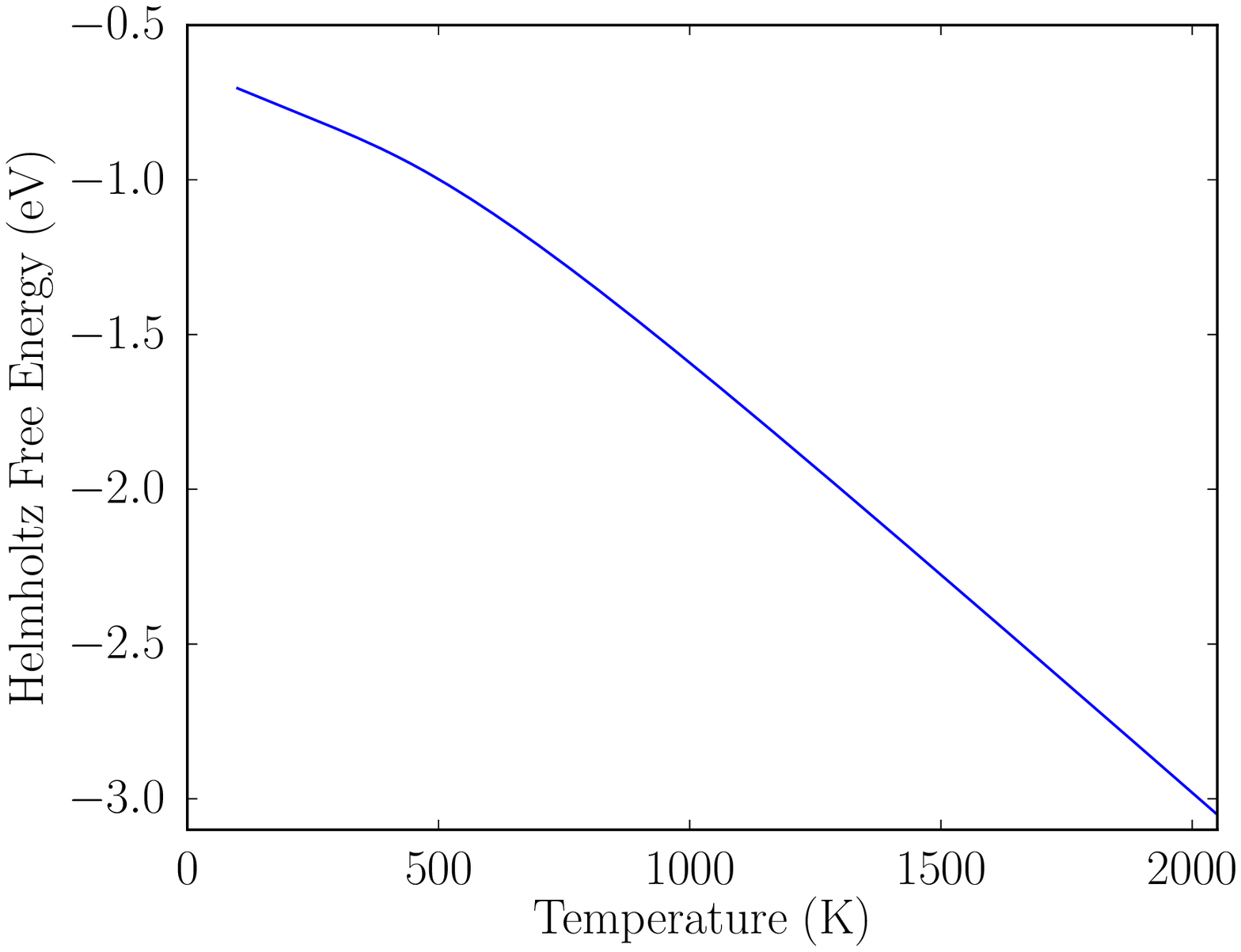} 
\caption{Temperature dependence of the Helmholtz free energy obtained with the Wang-Landau simulation approach for the model YSZ system with three charged defects (2 yttrium substitutions + 1 oxygen vacancy). The change in slope of the graph is indicative of the transition from an associated regime (dominated by the Coulombic interactions between the defects) to a dissociated regime (in which entropic effects dominate). }
\label{unfrozenFreeEn}
\end{center}
\end{figure}

One can now calculate the free energy of the model YSZ system as a function of temperature and defect density.
In Figure \ref{finiteSizeEffects} the density dependence is illustrated for isotherms at $T = 500\,$K, $T = 1000\,$K and $T = 2000\,$K. Also illustrated in 
this figure is the effect of the size of the supercell. 
In Figure \ref{unfrozenFreeEn} the temperature dependence of the Helmholtz free energy for the $1\%$ defect isochore 
is displayed. The slope of this curve is the negative entropy, and its change with temperature is indicative of the dissociation of the charge-neutral trimers, discussed in relation to Figure \ref{unfrozenIntEn}.

It is interesting to compare the relative efficiency of the WL and MMC routines in obtaining convergence of the
internal energy. In Table \ref{Tablet} we provide the number of energy evaluations 
(\textit{i.e.}, calculations of the total energy of the system) required, using each technique, to obtain convergence of the internal energy to an accuracy of $0.001$eV per defect. It is worth reiterating here that each simulation using the MMC technique 
relates to a 
single, fixed temperature, whereas the WL simulation technique provides a description over the full temperature range.
\begin{table}[ht]
  \begin{tabular}{| c | c | c |}
     \hline 
     Mole Fraction & \multicolumn{2}{c|}{Energy Evaluations} \\

     \;of Yttria &\; Wang-Landau\; &\;Metropolis\; \\
 \hline \hline 
    1\% & $1.1\;\mathrm{x}\;10^{8}$ &$9.0\;\mathrm{x}\;10^{8}$\\ \hline
     5\%  & $1.2\;\mathrm{x}\;10^{9}$ & $1.2\;\mathrm{x}\;10^{9}$\\ \hline
    10\% & $1.8\;\mathrm{x}\;10^{10}$&$1.6\;\mathrm{x}\;10^{9}$ \\
    \hline    
  \end{tabular}
  \caption{The number of energy evaluations required to obtain convergence of the internal energy to an accuracy of $0.001$eV per defect using the Wang-Landau and Metropolis Monte Carlo techniques.}
\label{Tablet}
\end{table}
One can observe that the relative efficiency of the two techniques is concentration dependent as one would expect. For low concentrations of defects the WL method requires fewer energy evaluations than MMC, so that WL obtains the  thermodynamics 
of the system at all temperatures at less computational cost than a single MMC simulation at fixed $T$. At higher concentrations the 
advantage is less spectacular; although with 10\% yttria a single WL simulation makes about  ten times more energy evaluations than a single MMC simulation, one  may require many more than ten MMC simulations to obtain the desired information. 
Furthermore, a number of strategies for improving the performance of the WL method exist \cite{Zhou2005,Morozov2007,Zhan2008}; a discussion of the merits of these strategies is beyond the scope of 
our current work, but a more rigorous analysis of the scaling behaviour of the methodology will be outlined in a forthcoming publication. 

\subsection{Comparison with an ideal solution}
\label{subsec:ideal-solution_comparison}
\begin{figure}[htp]
\begin{center}
\includegraphics[width=0.7\columnwidth]{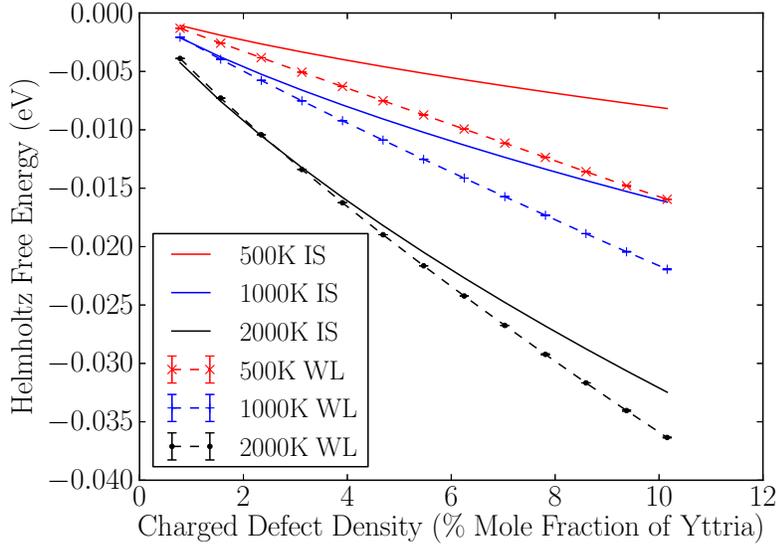} 
\caption{Comparison of the temperature and density dependence of the Helmholtz free energy of the system obtained with a Wang-Landau simulation approach (dashed curves, symbols) for the model YSZ system with an ideal-solution model (IS) of the defects (continuous curves). As expected the behaviour of the systems tends towards that of the ideal system, as the temperature increases. Energies are per site in the supercell.}
\label{agedIdealComp}
\end{center}
\end{figure}

One can readily observe from Figure \ref{agedIdealComp} that, as the temperature increases, the behaviour of the system tends towards that of an ideal solution  \cite{Callen1985}, which one can calculate using simple combinatorics for this lattice system, in which the only contribution to the free energy is the configurational entropy. This model thus takes the form 
\begin{equation}
 F_{\mathrm{ideal}}=\kB T\sum_{i=1}^{N_{c}} \left[\left(N_{i}-n_{i}\right)\ln\left(\frac{\left(N_{i}-n_{i}\right)}{N_{i}}\right)+n_{i}\ln\left(\frac{n_{i}}{N_{i}}\right)\right],
\end{equation}
where $n_{i}$ is the total number of a given charged species $i$, $N_{i}$ are the total number of lattice sites available to charged species $i$ and $N_{c}$ is the total number of charged species in the system.

The behaviour illustrated in Figure \ref{agedIdealComp} is what one would expect, given the behaviour already illustrated in Figures \ref{unfrozenIntEn} and \ref{unfrozenFreeEn} wherein, with increasing $T$, the system undergoes a transition from a regime in which the energetics (Coulombic interactions) dominate to one that is predominantly characterised by its entropy.
It is, however, important to point out that this behaviour is present only in the low-density regime. 
This also is to be expected as the ideal solution model is, by construction, exact in the limit of infinite dilution. 

\begin{figure}[htp]
\begin{center}
\includegraphics[width=0.7\columnwidth]{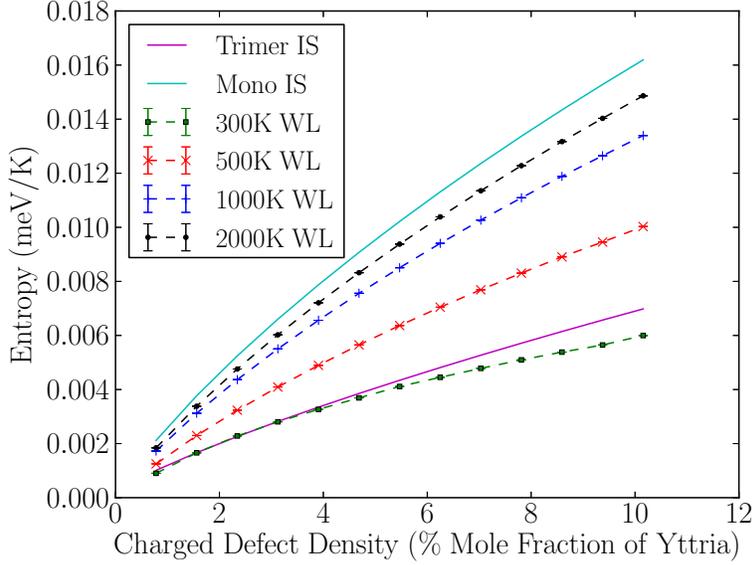} 
\caption{Comparison of the temperature and defect density dependence of the entropy obtained with a Wang-Landau simulation  (dashed curves, symbols) and two ideal solution models (solid curves), one of which describes the defects individually whilst the second describes the system as if it were comprised entirely of neutral trimers.  The change in the entropy  of the system with temperature illustrates the progression from the associated trimer to dissociated ion regimes. Entropies are per site in the supercell.}
\label{agedTrimerIdealComp}
\end{center}
\end{figure}

Since at the lower temperatures dimers and trimers are increasingly prominent, we can expect large departures from ideal behaviour.  To understand this in more detail it is helpful to study the behavour of the entropy, evaluated as $(U-F)/T$, which is plotted versus concentration at four different temperatures in Figure \ref{agedTrimerIdealComp}.
For comparison on this figure we have also plotted the entropy for  two extremes of ideal solution, in which the charges are $100\%$ associated as neutral trimers, which is the low temperature limit, and $100\%$ dissociated, which is the high temperature limit.
For the  ideal solution of trimers  the total number of possible microstates of the system is enumerated and the logarithm of this number gives the ideal entropy in units of $\kB$.
 At higher defect densities this model breaks down as the trimer-trimer interactions can no longer be neglected. This manifests itself by the formation of clusters of trimers. 
 At high temperatures, at which the free-energy is dominated by the individual charged defects, the ideal solution model of entropy is approached. 

\subsection{Comparison with Debye-H\"{u}ckel theory}
\label{subsec:D-H_comparison} 
For systems characterised by a low density of charged defects  an alternative approximation to the free energy can be obtained with  Debye-H\"{u}ckel (\mbox{DH}) theory \cite{Debye1923, Maier2004}. The DH contribution to the free energy is given by
\begin{equation}
 F_{\mathrm{DH}}=-\frac{\kB T\;\kappa^{3}\;V}{12\pi}
\end{equation}
and the internal energy by
\begin{equation}
 U_{\mathrm{DH}}=-\frac{\kB T\;\kappa^{3}\;V}{8\pi}\;,
\end{equation}
where $V$ is the volume of the system. The state-dependent parameter $\kappa$ is the reciprocal of the Debye length, and is defined as
\begin{equation}
 \kappa = \left(\frac{\sum_{i=1}^{N_{c}}n_{i}z_{i}^{2}}{\epsilon\;\kB T V}\right)^{\frac{1}{2}}\;;
\end{equation}
as before $N_{c}$ is the number of charged species in the system, $n_{i}$ is the total number of charged species $i$, of charge $z_{i}$; $\epsilon$ is the dielectric constant. 
Other approaches that should be more accurate than the ideal solution mode could be considered, such as the mean spherical approximation (MSA) \cite{Percus1964,Waisman1972,Blum1975}, but we do not consider these in the present work.

\begin{figure}[htp]
\begin{center}
\includegraphics[width=0.7\columnwidth]{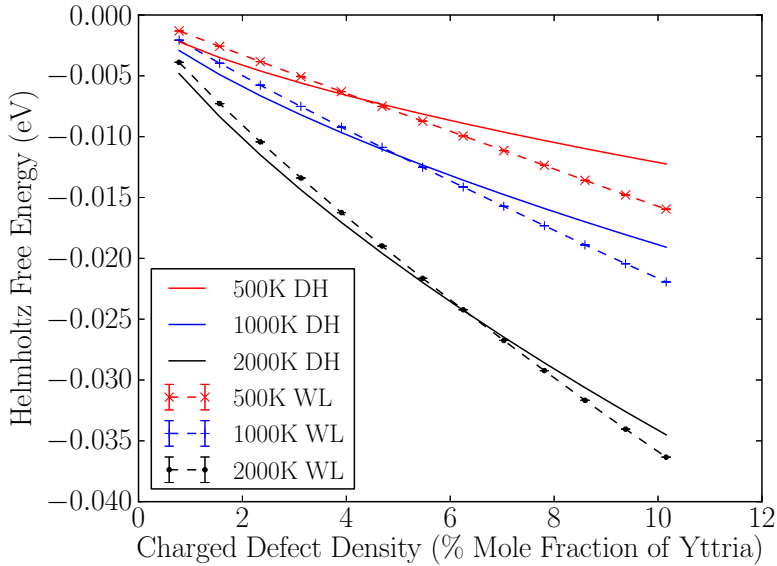} 
\caption{Comparison of the temperature and defect density dependence of the Helmholtz free energy obtained with the WL  approach (dashed curves, symbols) for the model YSZ system and the Debye-H\"{u}ckel theory (continuous curves). Energies are per site in the supercell.}
\label{agedDHComp}
\end{center}
\end{figure}

\begin{figure}[htp]
\begin{center}
\includegraphics[width=0.7\columnwidth]{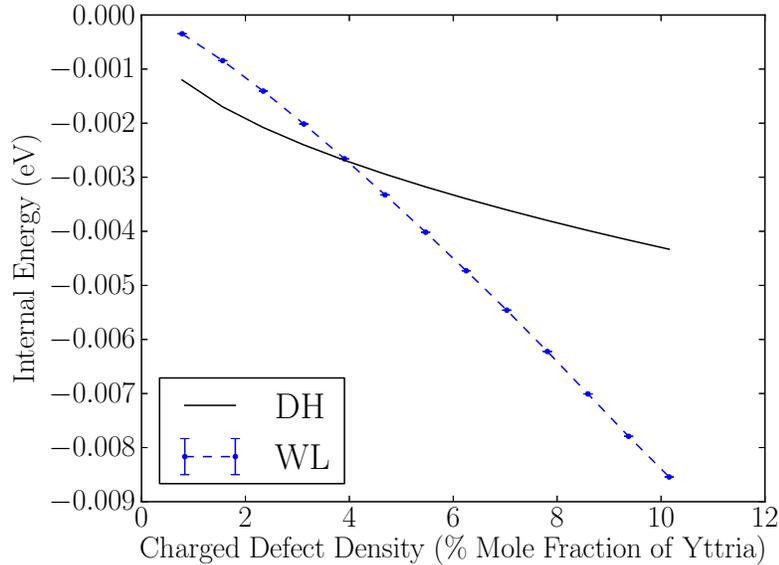} 
\caption{Comparison of the defect density dependence of the internal energy obtained with the Wang-Landau  approach (dashed curves, symbols)  and the Debye-H\"{u}ckel theory (continuous curves) at $1000\,$K. Energies are per site in the supercell.}
\label{DHintEn}
\end{center}
\end{figure}

 In Figure \ref{agedDHComp} we compare the \mbox{DH} free energy with that computed for our system with the WL approach.
There is reasonable agreement at higher temperatures, where both the \mbox{DH} description and the WL results tend to the ideal limit, however at low temperatures
the overall agreement between \mbox{DH} theory and  the WL simulations, which we consider to be the  `exact' results for this model system, becomes worse. 
On further analysis it becomes evident that this disagreement is due to the poor description of the internal energy of the system with the \mbox{DH} approach; this is illustrated in Figure \ref{DHintEn}. 
The poor quality of   the DH description can be understood in terms of the relatively high density of defects in our system which have a far from random distribution as they  tend  to lower the internal energy by forming dimers and trimers; this undermines the assumptions made in the formulation of \mbox{DH} theory, a mean-field theory that is strictly valid only for low charge densities.

\begin{figure}[htp]
\begin{center}
\includegraphics[width=0.7\columnwidth]{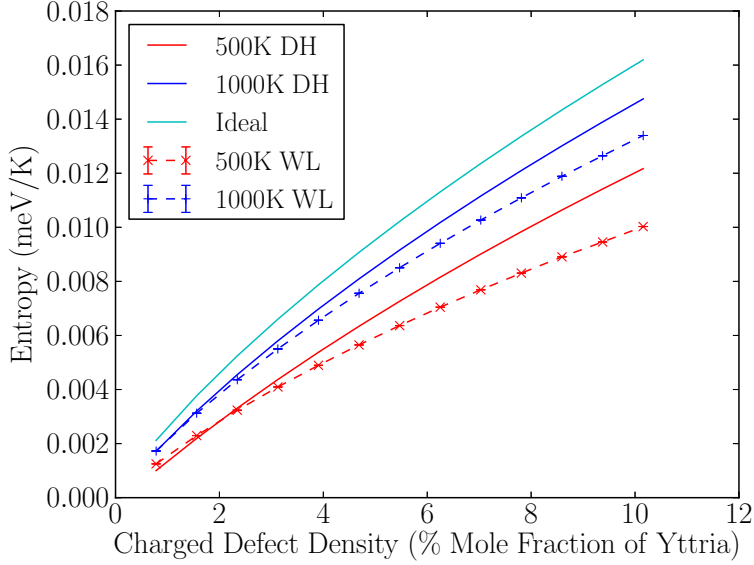} 
\caption{Comparison of the defect density dependence of the entropy (represented here as $S=-(F-U)/T$) obtained with the Wang-Landau simulation approach (dashed curves, symbols) for the model YSZ system and the Debye-H\"{u}ckel theory (continuous curves). Entropies are per site in the supercell.}
\label{DHentropy}
\end{center}
\end{figure}

\begin{figure}[htp]
\begin{center}
\includegraphics[width=0.7\columnwidth]{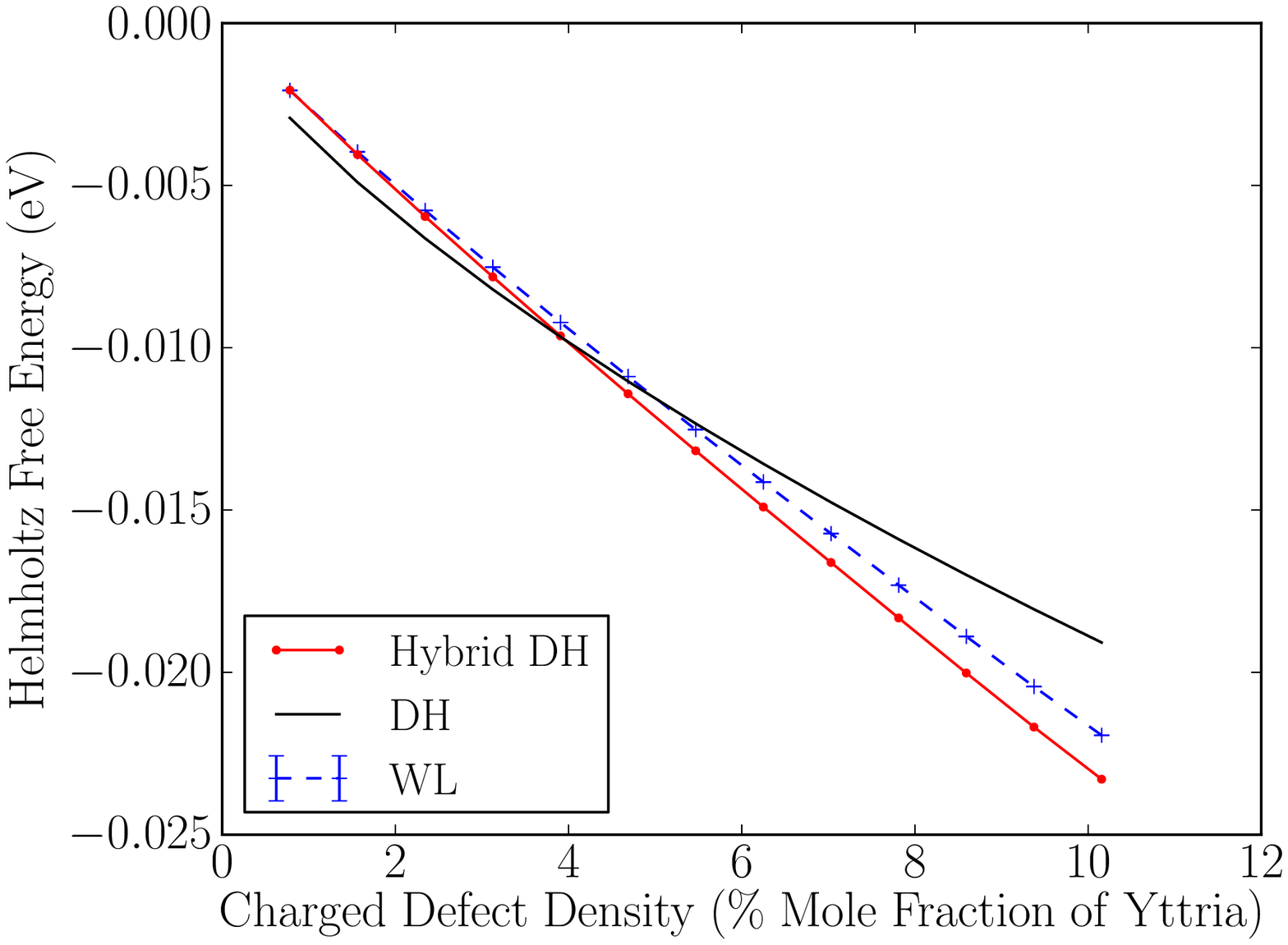} 
\caption{Comparison of the defect density dependence of the Helmholtz free energy at $1000\,$K obtained with a.) the full WL calculation (taken as the exact result for the model), b) the pure DH theory and c) a `hybrid'  DH free energy constructed from the  exact internal energy with the  DH entropy. Energies are per site in the supercell.}
\label{agedDHComp2}
\end{center}
\end{figure}

In contrast with the internal energy, one can observe from Figure \ref{DHentropy} that the entropic contribution obtained with the \mbox{DH} theory is in rather better agreement with the WL values, correcting some of the error made by the ideal approximation.  Thus given an MMC calculation of an accurate  internal energy at a single temperature one could make a rapid and reasonably good estimate of the free energy by using the DH value of entropy, as illustrated in Figure (\ref{agedDHComp2}).

\subsection{Regular solution Models}
\label{subsec:regular-solution_comparison}

It is useful to explore how an analytic  description of the free energy can be developed beyond the ideal solution model, which is only accurate at very low densities or high temperatures. In thermodynamic modelling of solutions, as in the CALPHAD approach \cite{Saunders1998,Lukas2007}, this is accomplished by adding an empirical polynomial in concentrations on the sublattices involved, which notionally represents the enthalpy of mixing. The leading term is a product  of the concentrations of the species, inspired by the Bragg-Williams model.
Thus the simple regular solution (RS) model is of the form
\begin{equation}
F_{\mathrm{RS}}=F_{\mathrm{ideal}}+\sum_{i}\sum_{j}x_{i}x_{j}L_{ij}
\end{equation}
where $x_{i}$ and $x_{j}$ are the concentrations of  two components and the $L_{ij}$ are adjustable parameters; in our implementation these are treated as temperature dependent.
\begin{figure}[htp]
\begin{center}
\includegraphics[width=0.7\columnwidth]{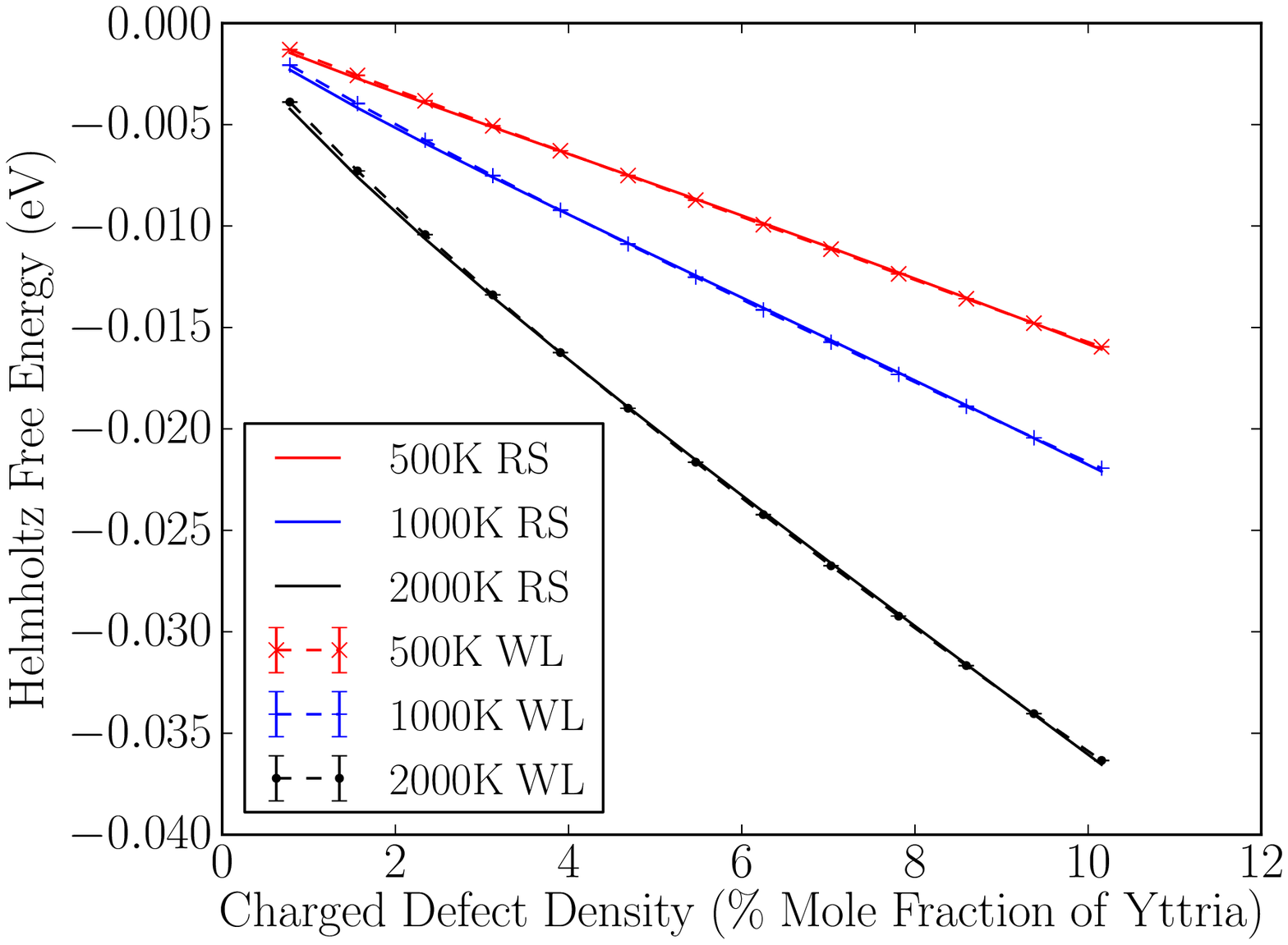} 
\caption{Free energies calculated using an empirical regular solution (RS) approach (continuous curves). 
Free energies obtained with our Wang-Landau simulation methodology (dashed curves, symbols) for the model YSZ system are
included for comparison (note that the error bars are of the order of magnitude of symbol size). Energies are per site in the supercell.}
\label{agedRegSolComp}
\end{center}
\end{figure}
One can observe from Figure \ref{agedRegSolComp} that a simple one parameter RS model of this type can be employed to provide a very good description of the thermodynamics of the system. 

Although for our simple system, a good representation of the thermodynamic properties can be achieved with the RS model, for more-complicated systems it may be beneficial to go a step further and use empirical extensions such as the Redlich-Kister (RK) \cite{Redlich1948} model, in which the parameters of the regular-solution model are taken to depend on the concentration of the species, \textit{i.e.},
\begin{equation}
 L_{ij}=\sum_{\nu=0}^{k}\left(x_{i}-x_{j}\right)^{\nu}A_{ij;\,{\nu}}
\end{equation}
where $A_{ij;\,{\nu}}$ are adjustable parameters.
This would enable  accurate simulations to be fitted analytically with arbitrary precision.


\section{Discussion and Conclusions}
\label{sec:Conclusions}


We have proposed a  methodology for the description of the thermodynamics of charged defects in crystals, which enables the free energy of a model of interacting point defects to be calculated accurately even at defect concentrations up to about $10\%$, as exhibited by complex oxides of practical interest, such as yttria stabilized zirconia. This is far above the range of validity of the ideal solution model that is commonly used in discussions of point defect equilibrium in ceramic materials.
In our approach, the temperature-independent Wang-Landau (WL) Monte Carlo method is used to calculate the free energy of the system at the atomistic level of an electrolyte lattice. As a test bed for this approach, we apply it to a very simple model of yttria stabilized zirconia, in which only the Coulomb interactions, dominant at all but nearest neighbour sites, are considered. 
The approach is much more efficient than conventional Metropolis Monte Carlo, since following a single, temperature-independent simulation one can obtain the Helmholtz free energy and entropy at \emph{all} temperatures of interest by simple quadrature over the density of states. 

We have shown that the free energy tends to  ideal solution behaviour in the high-temperature, low-density limit, as expected. 
A comparison of our WL calculations with the free energy obtained analytically with Debye-H\"uckel theory shows that the latter provides a reasonable description of the non-ideal entropy, but a poor description of the internal energy. 

Our resulting free energies as a function of temperature and defect concentration can be accurately fitted with a regular solution model using one temperature-dependent parameter. This suggests a promising route to providing input from physical models into the data needed for the calculation of phase diagrams. A more sophisticated model of the crystal and its defects could be parameterised from, e.g., DFT calculations, incorporating correctly the long-ranged Coulomb interaction, together with short ranged potentials or cluster models, and even the elastic interactions between defects. A free energy calculation using the methodology introduced here would then correctly include all the correlation between defect positions, and the energy and entropy therefrom that is neglected in the dilute or ideal solution model.

%
\noindent\textbf{Acknowledgments}\\
\noindent R. M. Horton was supported through a studentship in the Centre for Doctoral Training on Theory and Simulation of Materials at Imperial College funded by EPSRC under grant number EP/G036888/1. Additional funding to the Molecular Systems Engineering Group from the EPSRC (grants EP/E016340 and EP/J014958) is also gratefully acknowledged. We acknowledge support from the Thomas Young Centre under grant TYC-101.



\section*{References}

\begin{thebibliography}{29}
\bibitem{Goff1999}
Goff J P, Hayes W, Hull S, Hutchings M T and Clausen K N 1999
\newblock {\em Phys. Rev. B} \textbf{59} 14202--14219

\bibitem{Owens2000}
Owens B B 2000
\newblock {\em J. Power Sources} \textbf{90} 2--8

\bibitem{Wang1981}
Wang D Y, Park D S, Griffith J and Nowick A S 1981
\newblock {\em Solid State Ionics} \textbf{2} 95--105

\bibitem{Inaba1996}
Inaba H and Tagawa H 1996
\newblock {\em Solid State Ionics} \textbf{83} 1--16

\bibitem{Saunders1998}
Saunders N and Miodownik A. P. 1998
\newblock {\em {CALPHAD Calculation of Phase Diagrams}}.
\newblock Pergamon

\bibitem{Lukas2007}
Lukas H L, Fries S G and Sundman B 2007
\newblock {\em {Computational Thermodynamics: The Calphad Method}}.
\newblock Cambridge University Press

\bibitem{Wang2001}
Wang F and Landau D P 2001
\newblock {\em Phys. Rev. E} \textbf{64} 056101--056117

\bibitem{Wang2001a}
Wang F and Landau D P 2001
\newblock {\em Phys. Rev. Lett.} \textbf{86} 2050--2053

\bibitem{Redlich1948}
Redlich 0 and Kister A T 1948
\newblock {\em Ind. Eng. Chem.} \textbf{40} 345--348

\bibitem{Callen1985}
Callen H  1985
\newblock {\em {Thermodynamics and an Introduction to Thermostatistics}}.
\newblock Wiley

\bibitem{Debye1923}
Debye P and H\"{u}ckel E 1923
\newblock {\em Physikalische Zeitschrift} \textbf{24} 185--206

\bibitem{Frenkel2002}
Frenkel D and Smit B 2002
\newblock {\em {Understanding Molecular Simulation}}.
\newblock Academic Press 2nd edition

\bibitem{Ingel1986}
Ingel R P and {Lewis III} D 1986
\newblock {\em J. Am. Ceram. Soc.} \textbf{69} 325--332

\bibitem{Samara1990}
Samara G A 1990
\newblock {\em J. Appl. Phys.} \textbf{68} 4214

\bibitem{Metropolis}
 Metropolis N, Rosenbluth A W, Rosenbluth M N, Teller A H and Teller E 1953
\newblock {\em J. Chem. Phys.} \textbf{21} 1087--1092

\bibitem{Swendsen1987}
Swendsen R H and Wang J S 1987
\newblock {\em Phys. Rev. Lett.} \textbf{58} 86--88

\bibitem{Berg1992}
Berg B A and Neuhaus T 1992
\newblock {\em Phys. Rev. Lett.} \textbf{68} 9--12

\bibitem{Lee1993}
Lee J 1993 
\newblock {\em Phys. Rev. Lett.} \textbf{71} 211--214

\bibitem{Shell2002}
Shell M, Debenedetti P and Panagiotopoulos A 2002
\newblock {\em Phys. Rev. E} \textbf{66} 1--9

\bibitem{Eisenbach2009}
Eisenbach M, Zhou C G, Nicholson D M, Brown G, Larkin J and 
  Schulthess T C 2009
\newblock {\em Proceedings of the Conference on High Performance Computing
  Networking, Storage and Analysis - SC '09} 1--8

\bibitem{Rathore2004}
Rathore N, Yan Q and de~Pablo J J 2004
\newblock {\em J. Chem. Phys} \textbf{120} 5781--8

\bibitem{Shimoyama2011}
Shimoyama H, Nakamura H and Yonezawa Y 2011
\newblock {\em J. Chem. Phys} \textbf{134} 024109

\bibitem{Brown2011}
Brown G, Odbadrakh K, Nicholson D M and Eisenbach M 2011
\newblock {\em Phys. Rev E} \textbf{84} 065702--065706

\bibitem{Zhou2005}
Zhou C and Bhatt R 2005
\newblock {\em Phys. Rev. E} \textbf{72} 1--4

\bibitem{Morozov2007}
Morozov A and Lin S 2007
\newblock {\em Phys. Rev. E} \textbf{76} 026701

\bibitem{Zhan2008}
Zhan L 2008
\newblock {\em Comput. Phys. Commun.} \textbf{179} 339--344

\bibitem{Maier2004}
Maier J 2004 
\newblock {\em {Physical Chemistry of Ionic Materials}}.
\newblock Wiley

\bibitem{Percus1964}
Percus J K and Yevick G J 1964
\newblock {\em Phys. Rev.} \textbf{136} B290--B296

\bibitem{Waisman1972}
Waisman E 1972
\newblock {\em J. Chem. Phys} \textbf{56} 3093

\bibitem{Blum1975}
Blum L 1975
\newblock {\em Mol. Phys.} \textbf{30} 1529--1535
\end{thebibliography}
\end{document}